\documentclass[runningheads]{llncs}
\usepackage{graphicx}
\usepackage{subfigure}
\usepackage{paralist}
\usepackage{bm}
\usepackage{amsmath}
\usepackage{amsfonts}
\usepackage{booktabs}
\usepackage{multirow}
\usepackage{makecell}
\usepackage[misc]{ifsym}

\usepackage{array}
\begin{document}
%
\title{Modeling Long-Term and Short-Term Interests with Parallel Attentions for Session-based Recommendation}
%
\titlerunning{PAN}
%


\author{Jing Zhu \and Yanan Xu \and
Yanmin Zhu\textsuperscript{(\Letter)}}
%
%

\institute{Department of Computer Science and Engineering,\\
Shanghai Jiao Tong University, Shanghai, China\\
\email{\{sjtu\_zhujing,xuyanan2015,yzhu\}@sjtu.edu.cn}}
%
\maketitle              
\begin{abstract}
The aim of session-based recommendation is to predict the users' next clicked item, which is a challenging task due to the inherent uncertainty in user behaviors and anonymous implicit feedback information. A powerful session-based recommender can typically explore the users' evolving interests (i.e., a combination of his/her long-term and short-term interests). Recent advances in attention mechanisms have led to state-of-the-art methods for solving this task. However, there are two main drawbacks. First, most of the attention-based methods only simply utilize the last clicked item to represent the user's short-term interest ignoring the temporal information and behavior context, which may fail to capture the recent preference of users comprehensively. Second, current studies typically think long-term and short-term interests as equally important, but the importance of them should be user-specific. Therefore, we propose a novel Parallel Attention Network model (PAN) for Session-based Recommendation. Specifically, we propose a novel time-aware attention mechanism to learn user's short-term interest by taking into account the contextual information and temporal signals simultaneously. Besides,
we introduce a gated fusion method that adaptively integrates the user's long-term and short-term preferences to generate the hybrid interest representation. Experiments on the three real-world datasets show that PAN achieves obvious improvements than the state-of-the-art methods.

\end{abstract}
\section{Introduction} \label{1}
Recommender Systems play a significant role to provide personalized recommendations for different users in many application domains. Classical recommender systems typically utilize user historical interactions. In other words, user identity must be visible in each interaction record. However, in many application scenarios, the identities of users are unknown and the recommender system can employ only the user behavior history during ongoing sessions. To solve this problem, session-based recommendation~\cite{hidasi2015session,zhu2017next,rendle2010factorizing} is proposed to predict which item will be clicked by the user based on the sequence of the user's previous clicked items in the current session.

\begin{figure}
  \centering
  \includegraphics[scale=0.35]{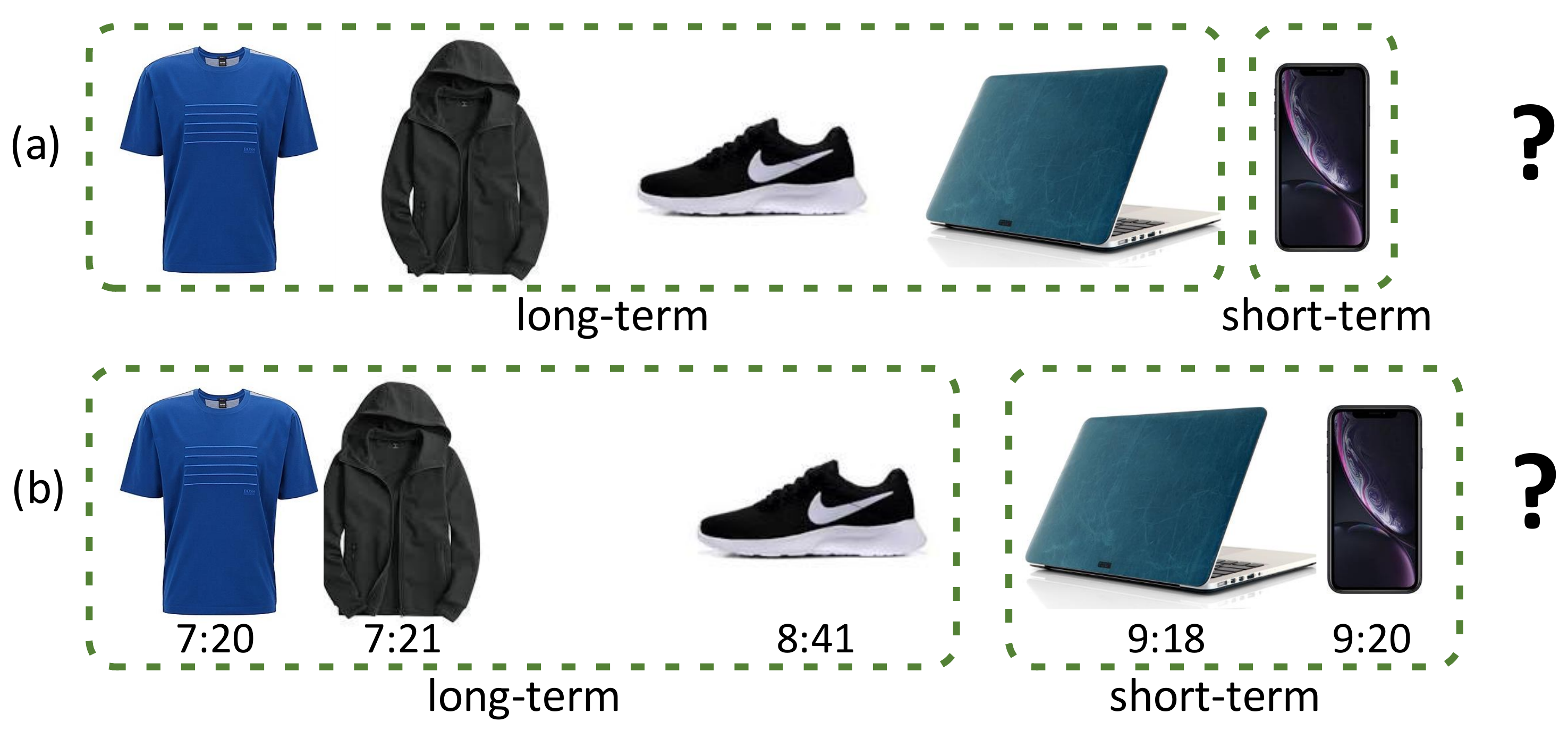}
  \caption{The main difference between the existing attention-based methods and our method. (a): The existing attention-based methods distinguish long-term interest and short-term interest, Nevertheless, they neglect temporal signals and behavior context when identifying short-term interest. (b): Our method not only distinguishes long-term interest and short-term user interest but also considers temporal signals and contextual information to identify short-term interest.} \label{fig2}
\end{figure}
Recently, the attention-based~\cite{hidasi2015session,liu2018stamp,song2019islf} methods have been largely developed in session-based recommendation. Compared with previous methods, attention-based models exhibit better capabilities to extract the user's interest (i.e., long-term interest or short-term interest) on user behavior data. Since Jannach et al.~\cite{jannach2015adaptation} confirmed that the long-term interest and short-term interest of users are both significant for recommendation, current attention-based methods~\cite{liu2018stamp,song2019islf,wu2019session} mainly model the user's long-term interest and short-term interest simultaneously. Although these studies have achieved state-of-the-art performance, there are also two main limitations, not thoroughly studies.


First, most existing attention-based methods~\cite{liu2018stamp,song2019islf,wu2019session} focus more on modeling long-term interest, simply regard the embedding vector of last clicked item as short-term interest. However, they neglect two important information: the time interval between actions, and the user's behavior context. On the one hand, user's short-term interest is changing over time, because past interest might disappear and new interest will come out. Intuitively, a user tends to have similar interests within a short time gap, while a large time gap may cause user's interests to drift. The time interval between actions is an important signal to indicate the change of user's interest. On the other hand, user's behavior sequence is much more complex. Only last clicked item cannot represent the short-term interest efficiently. For example, The last two clicked items of user $A$ and user $B$ are [$Apple\,Watch, MaBook\,Pro\,13$] and [$Surface\,Pro\,7, MacBook\,Pro\,13$], respectively. Existing attention-based methods will model their short-term interest into the same, because they both clicked $MacBook\,Pro13$ finally. Nevertheless, according to the former actions, we can infer that user $A$ may be interested in Apple series products, the current interest of user $B$ should be laptops. Their short-term interests are relatively different. In a nutshell, time interval and behavior context play a significant part in exploring the short-term interest of a user.
Second, the current methods~\cite{liu2018stamp,song2019islf,zhu2017next} typically treat long-term interest and short-term interest as equally important. However, in real scenarios, the significance of long-term and short-term interests should depend on the specific user. Different users may have various options for them. Therefore, how to fuse short-term and long-term interests is a key question. Simply merge, such as concatenate~\cite{li2017neural} and Hadamard product~\cite{liu2018stamp}, does not account for any interactions between long-term and short-term interests latent features, which is insufficient for learning the hybrid interest representation.

To solve the aforementioned limitations, we propose a model called Parallel Attention Network (PAN) for Session-based Recommendation. The model adopts two parallel attention layers: a short-term attention layer and a long-term attention layer, which are used to model user's short-term and long-term interests, respectively. Since the user's short-term interest is the current interest that changes over time, a novel time-aware attention mechanism is proposed to learn user's short-term interest, which explicitly considers contextual information and temporal signals simultaneously to adaptively weighted aggregate most recent clicked items. Conversely, the user's long-term interest in the session has a rare probability to change over time~\cite{song2019islf}, long-term attention layer extracts the user's long-term interest based on the whole sequential behavior characteristics. Then, we employ a gated fusion method to adaptively combine the long-term and short-term interests by taking into account the user's behavior characteristics. Eventually, the PAN makes recommendations based on the hybrid interest representation (i.e., the combination of long-term and short-term interests). Fig.~\ref{fig2} concludes the main difference between the existing methods and our method. The main contributions of the proposed model are as follows:
\begin{itemize}
    \item We propose a novel model PAN to enhance user's interest representation for session-based recommendation.
    \item We propose a novel time-aware attention mechanism, in which the attention weights are calculated by combining the embedding of last clicked item with the temporal signal (time interval).
    \item We introduce a gated fusion method to adaptively incorporate the long-term and short-term preferences according to the specific user. 
    \item The proposed model is evaluated on three benchmark datasets, Experimental results show that PAN achieves better performance than the existing state-of-the-art methods, Further studies demonstrate the proposed time-aware attention and gated fusion method play an important role.
\end{itemize}

The remainder of this paper is organized as follows: we first discuss the related work in the session-based recommender systems in Sect.~\ref{2}, and formulate the problem in Sect.~\ref{3}. Then we illustrate our proposed model in Sect.~\ref{4}. The experimental settings and results are presented in Sect.~\ref{5}. Finally, we conclude this paper in Sect.~\ref{6}.

\section{Related Work}\label{2}
In this section, we briefly review the related work on session-based recommendation from the following three aspects.

\subsection{Conventional Methods}
Matrix factorization~(MF)~\cite{koren2009matrix,rendle2009bpr,koren2015advances} is a classical solution to recommender systems. The idea of MF is to simultaneously map users and items into a continuous, latent and lower dimensional space. However, It has difficulty to apply in session-based recommendation because of the lack of user's information. A natural solution is item-based neighborhood methods~\cite{sarwar2001item}, in which item similarity matrix is pre-computed by the number of co-occurrence within the same session. Because the essence of session-based recommendation is a problem of sequence modeling, these methods are not suitable for the sequential problems. Then, a series of methods that are based on Markov chains~(MC) is proposed. MC predicts the next clicked item based on the previous clicked item by using sequence data. Shani et al.~\cite{rendle2010factorizing} utilize Markov decision processes~(MDPs) to solve the problem. FPMC~\cite{rendle2010factorizing} combines the power of MF and MC for next-item recommendation. However, the major limitation of Markov-chain-based models is that they cannot consider the whole sequence information because the state space will become enormous when taking into account all possible clicks.

\subsection{Recurrent-Network-based Methods}
Recurrent neural network (RNN) has been applied successfully in natural language processing area~\cite{mikolov2010recurrent,ma2018targeted}, therefore, a variety of RNN-based methods have emerged for session-based recommendation. Hidasi et al.~\cite{hidasi2015session} propose GRU4REC, which is a first work that applies RNN into session-based recommendation and achieves dramatical improvements than conventional models. They employ gated recurrent unit~(a variant of RNN) to model the sequential behavior data within the session and utilize a novel pair-wise loss function to optimize the model. Tan et al.~\cite{tan2016improved} enhance the performance of RNN-based methods by proposing two tricks, data augmentation and a method that can take temporal shifts in the behavior data into account. Zhu et al.~\cite{zhu2017next} proposed Time-LSTM that introduces time gates to model the time intervals between consecutive items to extract the users' long-term and short-term interests. Although RNN-based methods have achieved significant improvements for session-based recommendation, these methods suffer two limitations due to the RNN's drawbacks. First, both learning and inference processes are time-consuming due to its inherently sequential nature precluding parallelization~\cite{vaswani2017attention}. Second, RNN-based methods can model only simple transitions between consecutive items, for complex transitions between distant items, RNN is hard to capture~\cite{wu2019session}.
\subsection{Attention-Mechanism-based Methods}
Attention mechanism has been widely applied and shown to be effective in many tasks such as image caption~\cite{li2017image,xu2015show} and machine translation~\cite{luong2015effective,huang2016attention}. For session-based recommendation, Li et al.~\cite{li2017neural} propose NARM that is the first to employ attention mechanism for solving the task. They utilize an RNN and a vanilla attention method to capture the whole behavior sequence characteristic of users and the primary intention, respectively. Then to improve the session representation by considering the user's interests drift, Liu et al.~\cite{liu2018stamp} propose STAMP that can simultaneously extract the user's long-term and short-term interests. They extract the long-term interest of users by attention mechanism and utilize the embedding of last-clicked items as the short-term interests of users. A recent approach, ISLF~\cite{song2019islf} aims to capture the user's interest shift and latent factors simultaneously by an improved variational autoencoder and attention mechanism. However, these state-of-the-art attention-based methods ignore the importance of contextual information and temporal signals when learning short-term interest. In this study, we propose a time-aware attention mechanism that can take advantage of these two types of information.

\section{Problem Statement} \label{3}
The aim of session-based recommendation is to predict the next items that users will click, only based upon current sequential interaction data of anonymous users without any additional information. In this section, we formulate the session-based recommendation task as following.

In session-based recommendation, let $\mathbf{V} = \{ v_1, v_2, ..., v_{|V|} \}$ represents the set of all unique items that emerged in all sessions, called item dictionary. For each anonymous session, a click action sequence by unknown user can be denoted as a list $\mathbf{S}=[(s_1, t_1), (s_2, t_2), \ldots, (s_N, t_N)]$ ordered by timestamps, where $t_i$ is absolute time in seconds of the $i$-th click behavior, and $s_i$ is the $i$-th clicked item in the current session. $\mathbf{S}_n=[(s_1, t_1), (s_2, t_2), \ldots, (s_n, t_n)],1 \leq n \leq N$ denotes a prefix of the original interaction sequence $\mathbf{S}$ that truncated at $n$-th timestamp. Given session prefix $\mathbf{S_n}$, the aim of session-based recommender systems is to predict the user's next clicked item~(i.e., $s_{n+1}$). To be exact, the session-based recommendation task can be seen as learning a ranking model that ranks the candidate items to generate a recommendation list. Let $\hat{\mathbf{y}} = [\hat{y}_1, \hat{y}_2, ..., \hat{y}_{|V|}]$ denotes the output probability vector, where $\hat{y_i}$ corresponds to the probability of item $v_i$ being clicked at the next timestamp. The top-K items based on the scores of $\hat{\mathbf{y}}$ will be recommended for users.

\section{The Proposed Model} \label{4}
\begin{figure}[!t]
  \centering
  \includegraphics[scale=0.28]{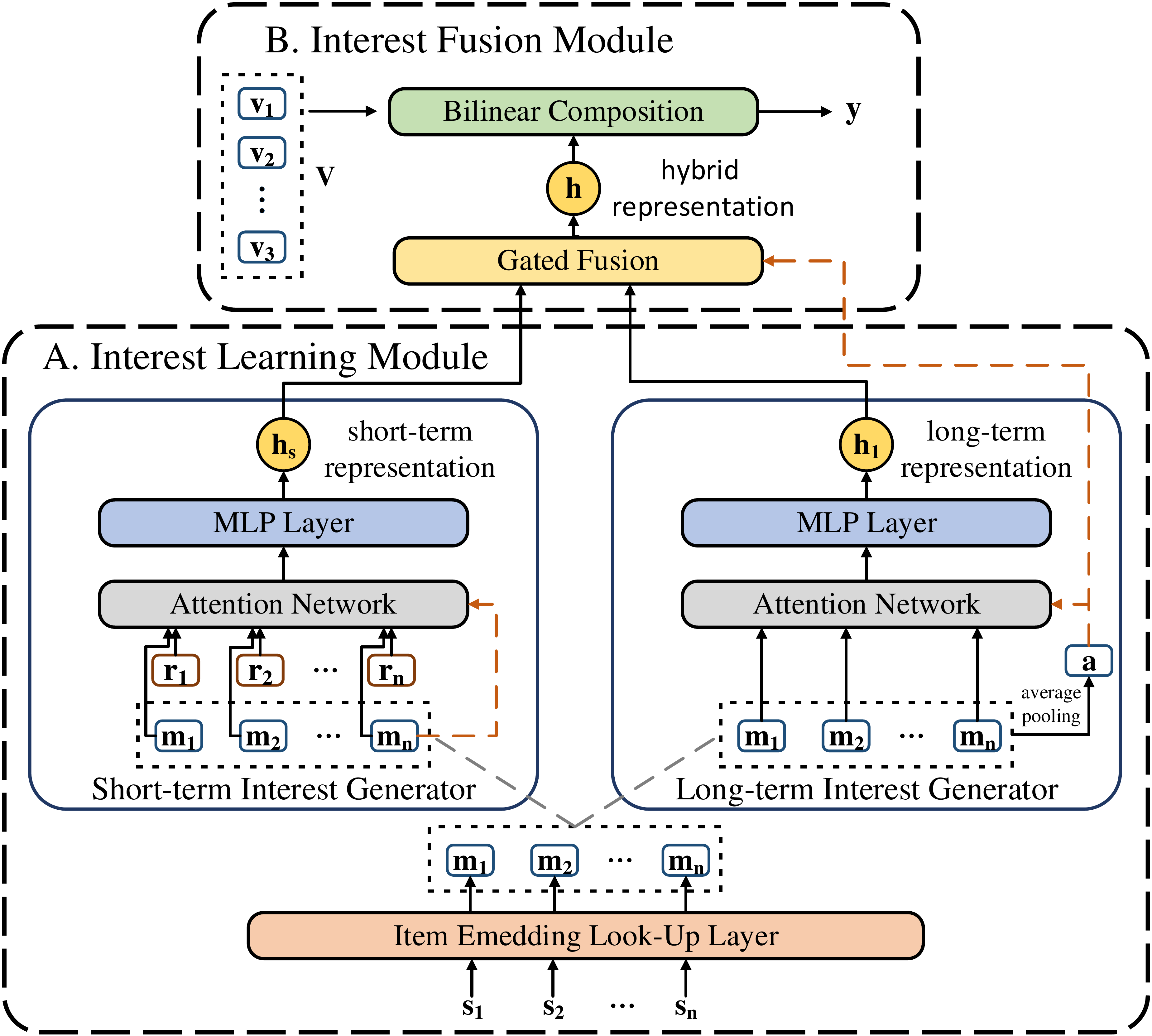}\\
  \caption{The architecture of PAN model.}
  \label{fig1}
\end{figure}

In this section, we will introduce parallel attention network (PAN) model in detail. As illustrated in Fig.~\ref{fig1}, there are two parts in our model: one is interest learning module; the other is interest fusion module.
\begin{enumerate}
     \item \textbf{Interest Learning Module.} The interest learning module consists of two components: short-term interest generator and long-term interest generator. The short-term interest generator utilizes a time-aware attention mechanism to learn short-term interest. The long-term interest generator extracts the long-term purpose within the session based on the behavior sequence representation.
    \item \textbf{Interest Fusion Module.} We integrate the short-term and long-term interests by a gated fusion method in interest fusion module. Then a bi-linear similarity function is utilized to compute recommendation score for each candidate item.
\end{enumerate}
\subsection{Interest Learning Module}
\subsubsection{Embedding Layer.}
Given a session prefix $\mathbf{S}_n=[s_1, s_2, \ldots, s_n]$, which is composed of item IDs and ordered by time. In embedding layer, The model utilizes an item embedding matrix $\mathbf{I} \in \mathbb{R}^{|V| * d}$ to map each item $s_i$ into the item vector $\mathbf{m}_i$, where $d$ is the dimensionality of item embedding vectors. Then we stack these item embedding vectors together to to obtain a set of latent item vectors $\mathbf{M} = [\mathbf{m}_1, \mathbf{m}_2, \ldots, \mathbf{m}_n]$.

\subsubsection{Short-term Attention Layer.}
As mentioned in Sect.~\ref{1}, the interest of online users is changing with time. To capture user's short-term interest, we think the user's last clicked item $\mathbf{m}_n$ can reflect some useful information. on the other hand, we think the time interval from last clicked item is an important signal to indicate the envolving of user’s interest. Therefore, a novel time-aware attention mechanism is proposed to learn the short-term interest from the current session, the attention mechanism considers contextual information and temporal signals simultaneously.  

First, we convert relative time signals to $d$-dimensional time interval embeddings. Let $\mathbf{r}_i$ denotes the relative time interval embedding vector of the $i$-th clicked item. Inspired by the work~\cite{vaswani2017attention,wang2019regularized},  $\mathbf{r}_i$ is computed as follows:
\begin{equation}
\begin{aligned}
&\mathbf{r_i}[2j]=\mathrm{sin}((t_n - t_i)/10000^{2j/d}) \\
&\mathbf{r_i}[2j+1]=\mathrm{cos}((t_n - t_i)/10000^{(2j+1)/d})
\end{aligned}
\end{equation}
where $t_n$ and $t_i$ are the absolute time in seconds of the last click and the $i$-th click. 

Next, we utilize a feed-forward network with one hidden layer to compute the short-term attention weight $\alpha_s^i$.
\begin{equation}
\begin{aligned}
&o_s^i = \mathbf{v}_s^\top\mathrm{tanh}(\mathbf{W}_s^q\mathbf{m}_n + \mathbf{W}_s^k\mathbf{m}_i + \mathbf{W}_s^r\mathbf{r}_i + \mathbf{b}_s^c) \\
&\alpha_s^i = \frac{exp(o_s^i)}{\sum_{i=1}^kexp(o_s^i)}
\end{aligned}
\end{equation}
where $\mathbf{W}_s^q, \mathbf{W}_s^k, \mathbf{W}_s^r \in \mathbb{R}^{d \times d}$ are weight matrices of the short-term attention network that covert $\mathbf{m}_n, \mathbf{m}_i, \mathbf{r}_i$ into a hidden layer, respectively; and $\mathbf{b}_s^c \in \mathbb{R}^d$ is the bias vector of the hidden layer; $\mathbf{v}_s \in \mathbb{R}^d$ is a weight vector to project the hidden layer output to a score $o_s^i$. Unlike a vanilla attention method that cannot be aware of the time interval between the target item and last-click item, we inject a time interval embedding vector $\mathbf{r}_i$ to model the effect of different time intervals. We choose tanh as the activation function in the hidden layer to enhance nonlinear capability. Then, we normalize the scores with a softmax function. At last, we can weighted sum the all item embedding vectors by the normalized scores to be the representation of the short-term interest of user.
\begin{equation}
    \mathbf{c}_s = \sum_{i=1}^n \alpha_s^i \mathbf{m}_i
\end{equation}
where $\mathbf{c}_s$ denotes user's short-term interest representation. 
\subsubsection{Long-term Attention Layer.}
As mentioned in Sect.~\ref{1}, the user's long-term interest within the current session is relatively difficult to change over time, so it is necessary to compute the user's attention on each clicked item. Previous studies~\cite{li2017neural,liu2018stamp} indicate that the user's overall behavior representation can provide useful information for capturing the user's long-term interest. Therefore, the long-term attention mechanism mainly considers overall behavior characteristics. Similar to STAMP~\cite{liu2018stamp}, We first utilize the average of all the item vectors within the session to summarize the whole session behaviors.

\begin{equation}
    \mathbf{a} = \frac{1}{n}\sum_{i=1}^n\mathbf{m}_i
\end{equation}
Where $\mathbf{a}$ denotes the whole sequential behavior representation. Then the user's long-term interest representation is computed as follows:
\begin{equation}
\begin{aligned}
&o_l^i = \mathbf{v}_l^\top\mathrm{tanh}(\mathbf{W}_l^q\mathbf{a} + \mathbf{W}_l^k\mathbf{m}_i + \mathbf{b}_l^c) \\
&\alpha_l^i = \frac{exp(o_l^i)}{\sum_{i=1}^kexp(o_l^i)} \\
&\mathbf{c}_l = \sum_{i=1}^n \alpha_l^i \mathbf{m}_i
\end{aligned}
\end{equation}
where $\mathbf{c}_l$ denotes the user's long-term interest representation; $\alpha_l^i$ denotes the long-term attention score to the $i$-th click item in the session; $\mathbf{W}_l^q, \mathbf{W}_l^k \in \mathbb{R}^{d \times d}$ are weight matrices, $\mathbf{b}_l^c$ is a bias vector, and $\mathbf{v}_l$ is a weight vector. 

\subsubsection{Multilayer Perceptron (MLP).}
It is beneficial to endow a high-level non-linearity abstraction to interest representation, which makes the model more expressive. Similar to STAMP~\cite{liu2018stamp}, we utilize two MLP networks that have one hidden layer to generate the hidden representation of long-term interest and short-term interest for considering interactions between different dimensions. The two MLP shown in Fig.~\ref{fig1} are pseudo-siamese network. In other words, they have the same structure but different learned parameters. To be specific, the operation on the short-term interest $\mathbf{c}_s$ is defined as:
\begin{equation}
    \mathbf{h}_s = f(\mathbf{W}_1^s \mathbf{c}_l + \mathbf{b}_1^s) \mathbf{W}_2^s  + \mathbf{b}_2^s
\end{equation}
where $\mathbf{W}_1^s, \mathbf{W}_2^s \in \mathbb{R}^{d \times d}$ are weight matrices; $f(\cdot)$ is a non-linear activation function (we found tanh has a better performance); and $\mathbf{h}_s$ denotes final short-term interest representation. The state vector $\mathbf{h}_l$ with regard to $\mathbf{c}_l$ can be computed similar as above.

\subsection{Interest Fusion Module}
The \emph{interest fusion module} adaptively combines the information from the short-term generator and long-term generator for recommendation. As mentioned in Sect.~\ref{1}, the relative significance between long-term interest and short-term interest is uncertain, depending on the specific user. Since user identification is unknown in session-based recommendation, we conjecture that the whole behavior representation $\mathbf{a}$ within the session may provide some user information, Inspired by the repeatNet~\cite{ren2019repeatnet}, we employ a gated fusion method, which can balance the significance of user's short-term interest and long-term interest to adaptively construct the hybrid interest representation:
\begin{equation}
    \mathbf{h} = \beta \mathbf{h}_s + (1 - \beta) \mathbf{h}_l
\end{equation}
where the gate $\beta$ is given by
\begin{equation}
    \beta = \sigma(\mathbf{W}_s \mathbf{h}_s + \mathbf{W}_l \mathbf{h}_l + \mathbf{W}_a \mathbf{a} + \mathbf{b})
\end{equation}
where $\mathbf{h}$ is the hybrid representation vector of user's interest, and $\sigma$ denotes sigmoid function. Similar to previous studies, then we exploit a bi-linear method to compute the recommendation score for each candidate item $v_i$:
\begin{equation}
    \hat{z}_i = emb_i^\top\mathbf{B}\,\mathbf{h}
\end{equation}
where $emb_i^T$ is the corresponding embedding vector to $v_i$, $\mathbf{B} \in \mathbb{R}^{d \times d}$ is a parameter matrix. We apply a softmax function to $\hat{\mathbf{z}} = [\hat{z}_1, \hat{z}_2, \ldots, \hat{z}_{|V|}]$ for obtaining the output vector $\hat{\mathbf{y}}$.
\begin{equation}
    \hat{\mathbf{y}} = \mathrm{softmax}(\hat{\mathbf{z}})
\end{equation}
where $\hat{\mathbf{y}} \in \mathbb{R}^{|V|}$ denotes the probability vector which is the output of PAN, and the utility of softmax function is to normalize the recommendation scores of all items into a probability distribution.
\subsection{Objective Function}
For any given session prefix $\mathbf{S}_n$, our model aims to maximize the prediction probability of the actual next clicked item within the session, which can be seen as a multi-classification problem. Thus, we define the loss function as the cross-entropy between prediction result $\mathbf{\hat{y}}$ and ground truth $\mathbf{y}$:
\begin{equation}
   \mathcal{L}(\hat{\mathbf{y}})=-\sum_{i=1}^{|V|} \mathrm{y}_{i} \log \left(\hat{\mathrm{y}}_{i}\right)+\left(1-\mathrm{y}_{i}\right) \log \left(1-\hat{\mathrm{y}}_{i}\right)
\end{equation}
Where $\mathrm{y}_i = 1$ indicates the user clicks this item. At last, we train our proposed PAN by utilizing the Back-Propagation~(BP) algorithm.

\section{Experiments} \label{5}
In this section, we perform extensive experiments on two benchmark datasets to evaluate our proposed PAN. We aim to answer the following research questions:

\textbf{RQ1} Can our proposed approach perform better than other competitive methods?

\textbf{RQ2} Are the key components in PAN (i.e., time-aware attention layer, gated fusion) useful for improving recommendation results?

\textbf{RQ3} How does session length affect the recommendation performance of our approach?
\subsection{Experimental Settings}
\subsubsection{Datasets.}
To evaluate the effectiveness and efficiency of our proposed PAN, we experiment with two benchmark datasets: YOOCHOSE and DIGENTICA. YOOCHOOSE is from RecSys'15 Challenge\footnote{http://2015.recsyschallenge.com/challege.html}, which contains a collection of sessions that composed of user clicks from a retailer website. DIGENTICA is from CIKM Cup 2016\footnote{http://cikm2016.cs.iupui.edu/cikm-cup}, where we only use the transaction data in this paper.

For a fair comparison, following~\cite{hidasi2015session,li2017neural}, The sessions with length one, items with less than five occurrences in the datasets, and items in the test set which do not appear in the training set are all filtered out. Similar to~\cite{tan2016improved}, we split each sequence to augment the data. To be specific, for the input Session $\mathbf{S} = [s_1, s_2, \ldots, s_n]$, we divide a series of sequences and corresponding labels ($[s_1]$, $s_2$), ($[s_1, s_2]$, $s_3$), $\ldots$~, ($[s_1, s_2, \dots, s_{n-1}]$, $s_n$) for two datasets, which has demonstrated better effectiveness in~\cite{tan2016improved}. For YOOCHOOSE dataset, The recent fractions 1/64 and 1/4 of training sequences are used to train the proposed model. The reasons for this action are that the YOOCHOOSE dataset is fairly enormous that has a high training time cost, and the experimental results of~\cite{tan2016improved} prove that utilize the recent fraction of training sequences has better performance than training on the whole data. The statistics of datasets that have been preprocessed are summarized in Table~\ref{dataset}.
\begin{table}[h]
  \centering
  \caption{Statistics of datasetss}
  \label{dataset}
  \setlength{\tabcolsep}{2mm}
    \begin{tabular}{cccc}
    \toprule
    Dataset & YOOCHOSE 1/64 & YOOCHOSE 1/4 & DIGINETICA \\
    \midrule
    \# train & 369,859 & 5,917,746 & 719,470 \\
    \# test & 55,898 & 55,898 & 60,858 \\
    \# clicks & 557,248 & 8,326,407 & 982,961 \\
    \# items & 17,745 & 29,618 & 43,097 \\
    avg.length & 6.16  & 5.71  & 5.12 \\
    \toprule
    \end{tabular}%
  \label{tab:addlabel}%
\end{table}%
\subsubsection{Baselines.}
For the aim of evaluating the performance of PAN, we compare it with the following baselines:
\begin{itemize}
\item \textbf{POP}: A non-personalized recommendation method, which only recommends items that have high popularity. Popularity is assessed by the number of occurrences in the training set.
\item \textbf{Item-KNN}~\cite{sarwar2001item}: An item-to-item model, which makes recommendations based on the cosine similarity between two items.
\item \textbf{BPR-MF}~\cite{rendle2009bpr}: A classical matrix factorization method, which utilizes a stochastic gradient descent algorithm to optimize a pair-wise loss function.
\item \textbf{FPMC}~\cite{rendle2010factorizing}: A classic hybrid model combining Markov chain and matrix factorization.
\item \textbf{GRU4REC}~\cite{hidasi2015session}: An RNN-based model, which employs a pair-wise loss function and a session-parallel mini-batch training process for training the model.
\item \textbf{Time-LSTM}~\cite{zhu2017next}: An improved RNN model, which proposes time gates to model the time intervals between consecutive items to capture user's long-term and short-term interests.
\item \textbf{NARM}~\cite{li2017neural}: An attention-based model, which utilizes an RNN and a vanilla attention method to capture the whole behavior sequence characteristic of users and the primary intention, respectively.
\item \textbf{STAMP}~\cite{liu2018stamp}: A novel memory model, which extract the long-term interest of users by an attention mechanism and utilize the embedding of last-clicked items as the short-term interest.
\item \textbf{ISLF}~\cite{song2019islf}: A state-of-the-art model for session-based recommendation, which employs an improved variational autoencoder with an attention mechanism to capture the user's dynamic interest and latent factors simultaneously.
\end{itemize}

\subsubsection{Evalution Metrics.}
We adopt the following two evaluation metrics that are widely used in session-based recommendation to evaluate the performance of all models.

\textbf{Recall@20}: Recall is a widely used evaluation metric in session-based recommendation, which cannot take the rank of items into account. Recall@K indicates the proportion of test cases, which have the actual next clicked items among the top-K ranking list. 
\begin{equation}
    \mathrm{Recall}@K = \frac{n_{hit}}{N}
\end{equation}
where $n_{hit}$ is the number of testing samples which have desired items in recommendation lists, $N$ is the number of the whole testing samples.

\textbf{MRR@20}: MRR is the average of reciprocal ranks of the actual next clicked item. MRR@20 will be 0 if the rank is above 20.
\begin{equation}
    \mathrm{MRR}@K = \frac{1}{N}\sum_{t \in G} \frac{1}{Rank(t)}
\end{equation}
where $G$ is the ground-truth set of the test cases. The MRR considers the order of the recommendation list, if the actual next clicked item is at the top of the recommendation list, the MRR value will be larger.

\subsubsection{Parameter Settings.}
We utilize a validation set that is randomly selected 10\% in the training data to tune the hyper-parameters. The dimensionality of latent vectors is searched in $\{16, 32, 64, 128\}$, and 128 is optimal. We choose Adam as the model learning algorithm to optimize these parameters, where the mini-batch size is searched in $\{64, 128, 256, 512\}$, and sets 128 finally. We initialize the learning rate to 0.001, and it will decay by 0.1 after every ten epochs. Following the previous method~\cite{liu2018stamp}, We randomly initialize all weight matrices with a Gaussian distribution $N(0, 0.05^2)$. For the item embedding matrix, we randomly initialized it using a Gaussian distribution $N(0, 0.002^2)$. We use dropout~\cite{srivastava2014dropout} with drop ration $\rho=0.5$. The model is written in Tensorflow and trained on an NVIDIA TITAN Xp GPU.
\subsection{Performance Comparison (RQ1)}
The overall performances of all contrast methods based on three datasets are shown in Table~\ref{overallPerf}. Please note that, as in~\cite{li2017neural}, We have not show the performance of FPMC on Yoochoose 1/4 because of the limited memory space. From the table, we make the following observations from the results.

The performance of conventional personalized methods such as BPR-MF and FPMC significantly outperform the naive POP method. This proves that modeling personalized preference is effective for improving the recommendation performance. Compared with the deep-learning-based methods, we can see that extracting co-occurrence popularity or employing a simple first-order transition probability matrix may not be sufficient for session-based recommendation. This indicates the importance of considering the current whole behavior context for recommendations.
\begin{table}[t]
  \centering
  \caption{The overall performance over three datasets}
    \label{overallPerf}
    \begin{tabular}{ccccccc}
    \toprule
    \multirow{2}[0]{*}{method} & \multicolumn{2}{c}{YOOCHOSE 1/64} & \multicolumn{2}{c}{YOOCHOSE 1/4} & \multicolumn{2}{c}{DIGINETICA} \\
    \cmidrule(r){2-3} \cmidrule(r){4-5} \cmidrule(r){6-7}
          & Recall@20 & MRR@20 & Recall@20 & MRR@20 & Recall@20 & MRR@20 \\
    \midrule
    POP & 6.71  & 1.65  & 1.33  & 0.30  & 0.89  & 0.20  \\
    Item-KNN & 51.60  & 21.82  & 52.31  & 21.70  & 35.75  & 11.57  \\
    BPR-MF & 31.31  & 12.08  & 3.40  & 1.57  & 5.24  & 1.98  \\
    FPMC  & 45.62  & 15.01  & -     & -     & 26.53  & 6.95  \\
    \midrule
    GRU4REC & 60.64  & 22.89  & 59.53  & 22.60  & 29.45  & 8.33  \\
    Time-LSTM & 67.78  & 28.12  & 69.14  & 28.96  & 47.13  & 15.22  \\
    NARM  & 68.32  & 28.63  & 69.73  & 29.23  & 49.70  & 16.17  \\
    STAMP & 68.74  & 29.67  & 70.44  & 30.00  & 45.64  & 14.32  \\
    ISLF  & 69.32  & \textbf{33.58} & 71.02  & 32.98  & 49.35  & 16.41  \\
    PAN & \textbf{70.36} & 31.97 & \textbf{71.58} & \textbf{33.04} & \textbf{50.69} & \textbf{16.53} \\
    \toprule
    \end{tabular}
  \label{tab:addlabel}%
\end{table}%

As can be seen, deep-learning-based methods outperform all traditional methods. This demonstrates the effectiveness of deep learning technology for session-based recommendation. Time-LSTM utilizes a time gate to extract the user's dynamic interest and improves the performances of GRU4REC, which indicates that temporal signals are helpful to capture user's shifty interest. PAN significantly outperforms all baseline methods. Generally, PAN obtains improvements over the best baseline ISLF of 1.50\%, 0.79\%, and 2.72\% in Recall@20 on the three datasets, respectively. Although both Time-LSTM and PAN taking temporal signals into account, we notice that PAN achieves better performance than Time-LSTM on both datasets. One reason might be that RNN is only adept at capturing the simple transitions between consecutive items. For other complex transitions between distant items, RNN is insufficient to model. However, attention mechanism can capture complex transitions within the entire session sequence. As to other baseline methods that also consider user's long- and short-term interests simultaneous~(i.e., STAMP and ISLF), we find that PAN outperforms STAMP and ISLF on all datasets. This is because PAN learns user’s current interest by taking into account the contextual information and temporal signals simultaneously. Conversely, STAMP and ISLF simply utilize the last clicked item, which may not be sufficient. 

\subsection{Effects of Key Components (RQ2)}
\subsubsection{Impact of Time-aware Attention.}
In order to prove the effectiveness of the proposed time-aware attention mechanism, we introduce the following variant methods of PAN. (1) PAN-l excludes the short-term interest. (2) PAN-v utilizes a vanilla attention mechanism that excludes the time interval information to learn short-term interest.
\begin{table}[htbp]
  \centering
  \caption{Impact of Time-aware Attention} \label{attPerfm}
    \begin{tabular}{ccccccc}
    \toprule
    \multirow{2}[0]{*}{method} & \multicolumn{2}{c}{YOOCHOSE 1/64} & \multicolumn{2}{c}{YOOCHOSE 1/4} & \multicolumn{2}{c}{DIGINETICA} \\
    \cmidrule(r){2-3} \cmidrule(r){4-5} \cmidrule(r){6-7}
          & Recall@20 & MRR@20 & Recall@20 & MRR@20 & Recall@20 & MRR@20 \\
    \midrule
    PAN-l & 67.21 & 28.03 & 68.88 & 31.13 & 46.97 & 13.06 \\
    PAN-v & 69.72 & 30.72 & 71.33 & 32.99 & 49.02 & 15.20 \\
    PAN & \textbf{70.36} & \textbf{31.97} & \textbf{71.58} & \textbf{33.04} & \textbf{50.69} & \textbf{16.53} \\
    \toprule
    \end{tabular}
  \label{tab:addlabel}%
\end{table}%
Table~\ref{attPerfm} shows the experimental results of the Recall@20 and MRR@20 metrics on all datasets. We can see that PAN-v outperform PAN-l for all datasets. For example, for YOOCHOSE 1/64, the PAN-v improves Recall@20 and MRR@20 by 3.96\% and 2.31\%, which indicates that the effectiveness of considering the short-term interest representation when capturing the user's overall representation. We can also see that PAN outperforms PAN-v for all datasets, such as PAN improves Recall@20 and MRR@20 by 0.57\% and 3.90\% on YOOCHOSE 1/64, which proves that the temporal signals are conducive to capturing the user’s short-term interest.

\subsubsection{Impact of Fusion Operations.}
To evaluate the effectiveness of the gated fusion, we consider the variants of PAN that utilize three other fusion operations, i.e., concatenation, average pooling, and Hadamard product. 
\begin{table}[htbp]
  \centering
  \caption{Impact of Gated Fusion} \label{fusionPerfm}
    \begin{tabular}{ccccccc}
    \toprule
    \multirow{2}[0]{*}{method} & \multicolumn{2}{c}{YOOCHOSE 1/64} &  \multicolumn{2}{c}{DIGINETICA} \\
    \cmidrule(r){2-3} \cmidrule(r){4-5}
          & Recall@20 & MRR@20 & Recall@20 & MRR@20 \\
    \midrule
    Average pooling & 69.77 & 31.58 & 49.99 & 16.02 \\
    Hadamard product & 70.18 & 31.79 & 50.13 & 16.37 \\
    Concatenation & 70.11 & 31.82 & 50.35 & 16.37 \\
    Gated fusion & \textbf{70.36} & \textbf{31.97} & \textbf{50.69} & \textbf{16.53} \\
    \toprule
    \end{tabular}
  \label{tab:addlabel}%
\end{table}%
As shown in Table~\ref{fusionPerfm}, the introduced gated fusion method achieves better results than the other three fusion operations on two datasets. This demonstrated that the gated fusion method makes an improvement in learning the latent interactions between user's long-term and short-term interests. In most cases, PAN with Hadamard product and PAN with concatenation operation perform nearly, and both exceed the average pooling method. This indicates that Hadamard product and concatenation operation have advantages than average pooling when learning interactions between the user's long-term and short-term interests latent features.
\begin{figure}[t]
\centering
\subfigure[YOOCHOSE-Recall]{
\includegraphics[width=5.5cm]{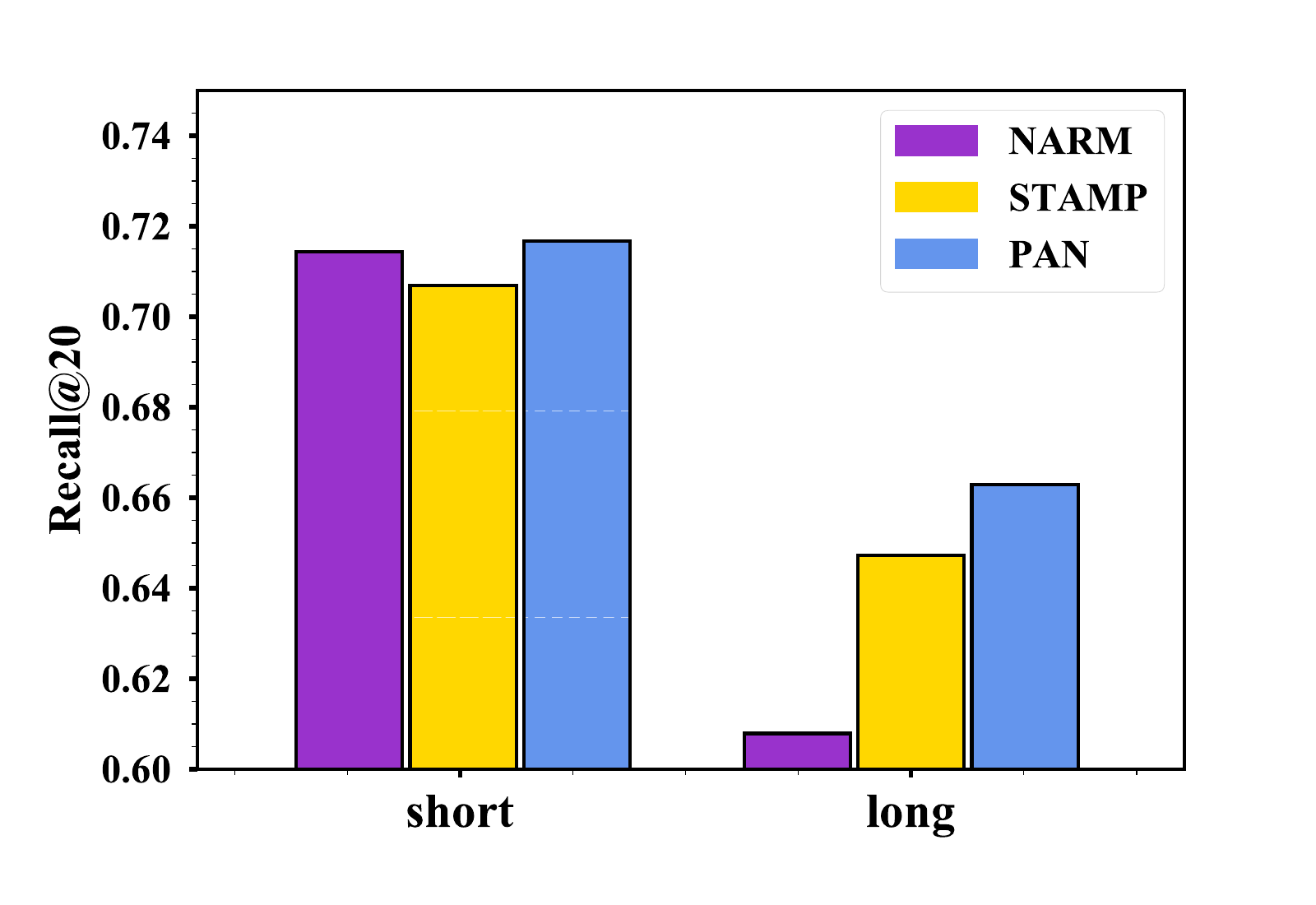}
}
\quad
\subfigure[YOOCHOSE-MRR]{
\includegraphics[width=5.5cm]{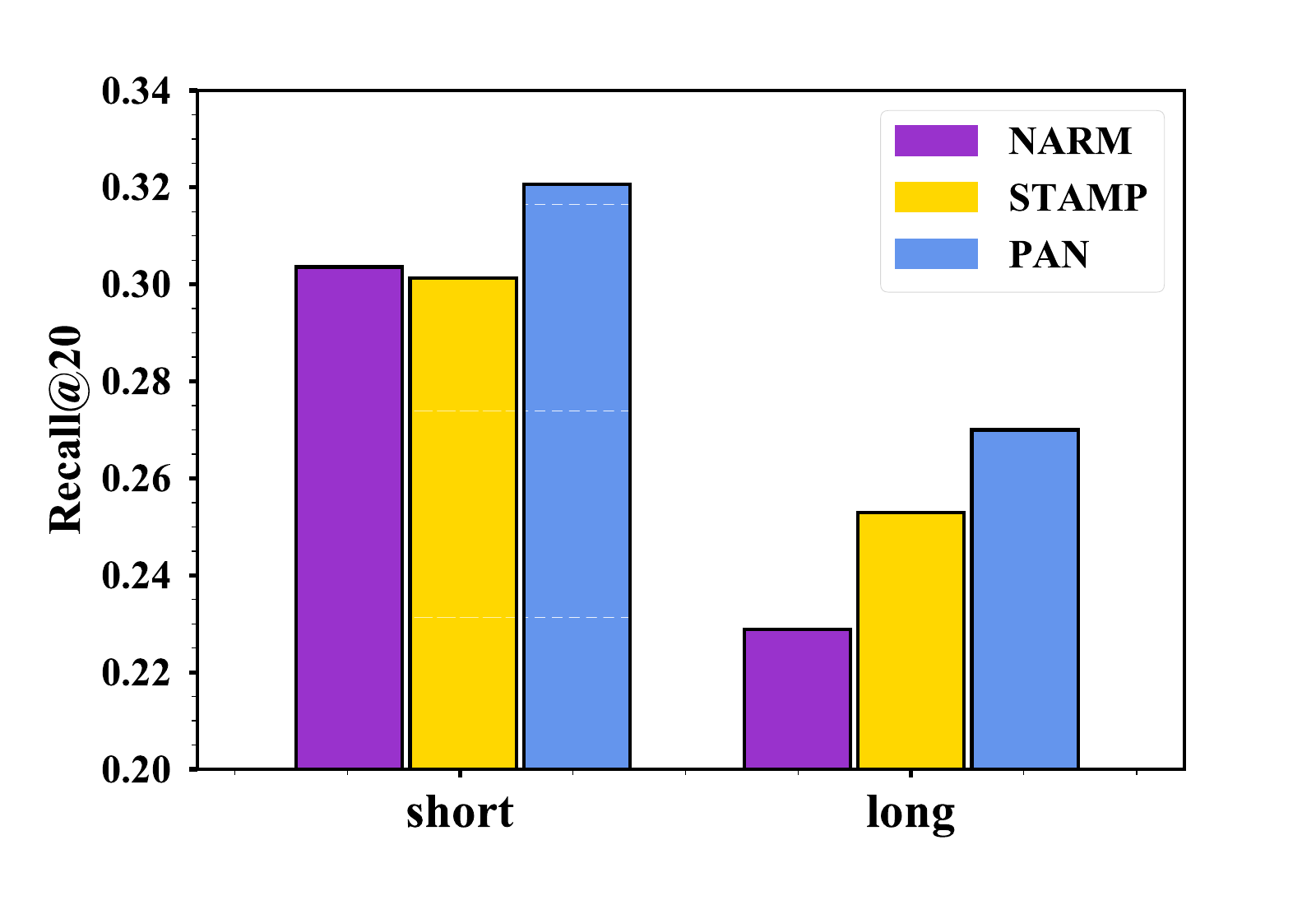}
}
\quad
\subfigure[Diginetica-Recall]{
\includegraphics[width=5.5cm]{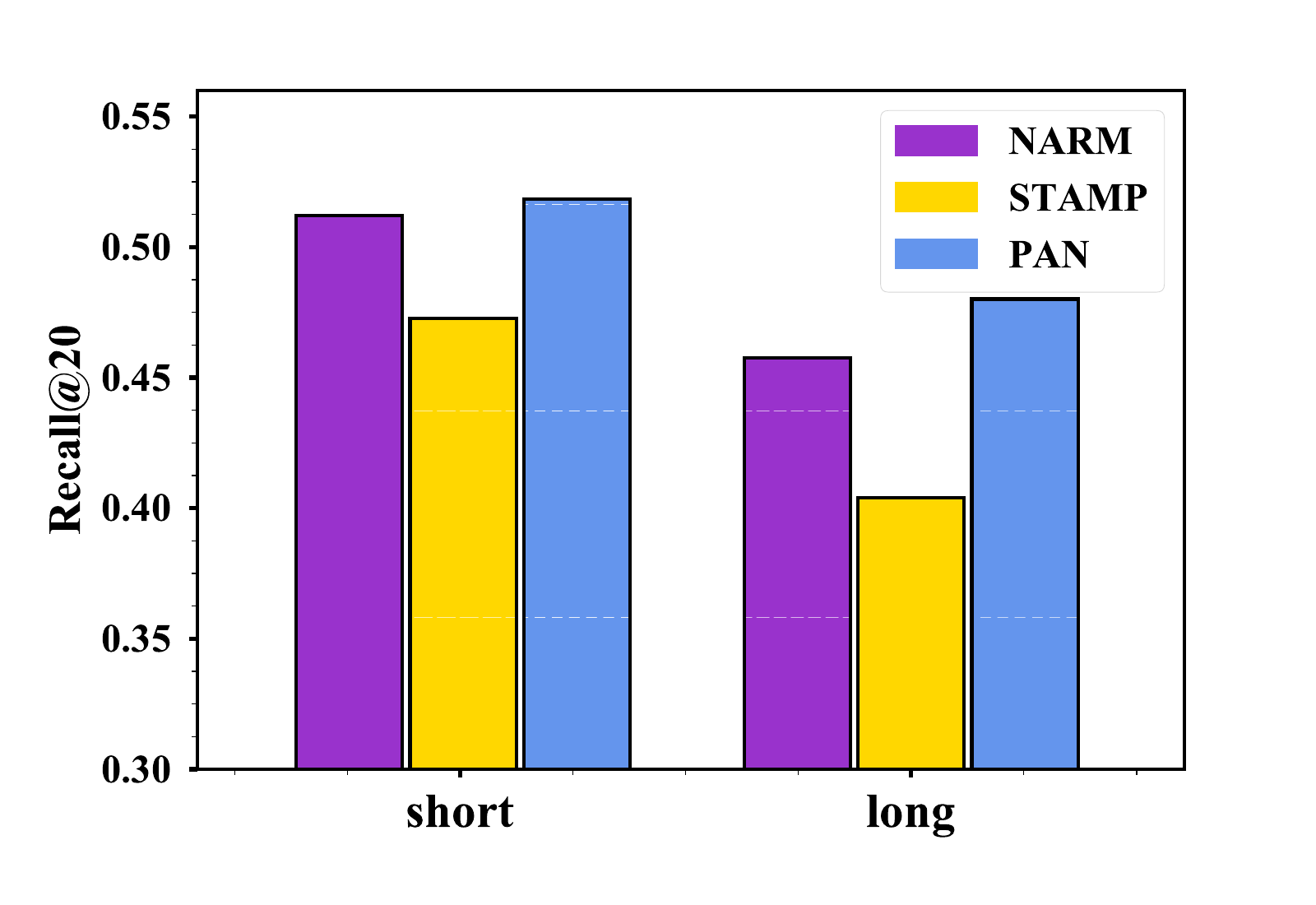}
}
\quad
\subfigure[Diginetica-MRR]{
\includegraphics[width=5.5cm]{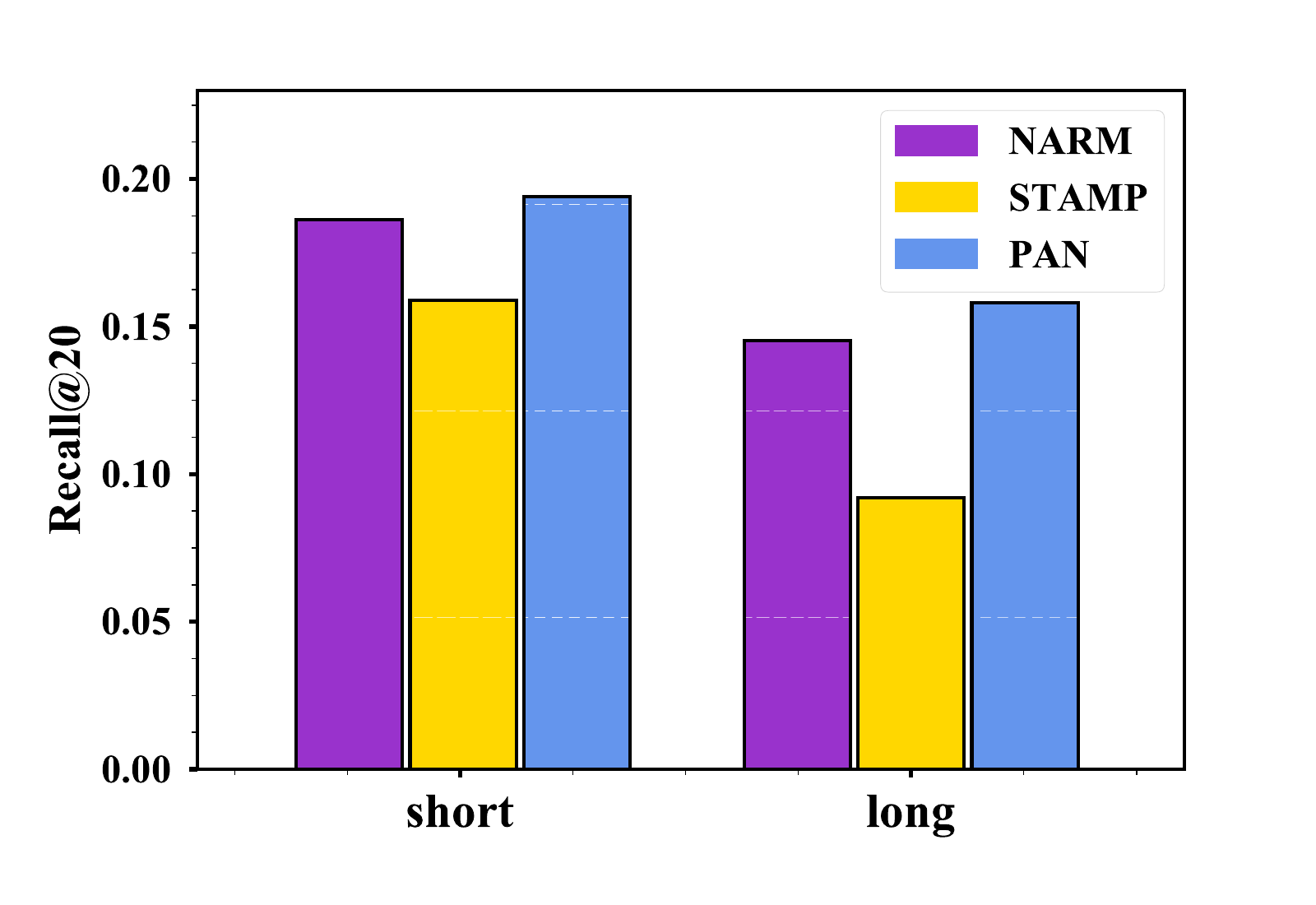}
}
\caption{Recall@20 and MRR@20 on different session length} \label{length_fig}
\end{figure}

\subsection{Influence of Different Session Lengths (RQ3)}

We further analyze the impact of different session lengths on different methods. For a comparison, following~\cite{liu2018stamp}, Yoochoose 1/64 and Diginetica, these two datasets are both divided into two groups in terms of the length of sessions. Since the average length of sessions is almost five, sessions with length greater than five are seen as long sessions, the remainder is called short sessions. The percentages of sessions belong to short sessions and long sessions are 70.1\% and 29.9\% on the YOOCHOSE 1/64, AND 76.4\% and 23.6\% on the Diginetica. We use Recall@20 and MRR@20 to compare the performance.

Fig.~\ref{length_fig} shows the comparison results with different groups and methods. We can observe that all methods obtain worse performance in long groups than short groups, which reveals the difficulty of handling long sequences in session-based recommendation. The performance of NARM changes greatly. The reason for this phenomenon may be that NARM neglects the user's short-term interest. Similar to NARM, STAMP achieves better performance in the short sessions than long sessions. STAMP explains this difference based on repetitive click actions. The reason is that duplicate items might be neglected due to employing attention method to extract the general interests of users. PAN is relatively stable than NARM and STAMP.

\section{Conclusion} \label{6}
In this work, we propose a novel parallel attention network (PAN) for session-based recommendation to modeling user's short-term and long-term interests. We observe that user's short-term interest has great relations to temporal signal and behavior context, thus we propose a novel time-aware attention mechanism to learning the short-term interest. We further introduce a gated fusion method to adaptively incorporate the long-term and short-term interests according to the specific behavior context. Extensive experimental analysis on three real-world datasets shows that our proposed model PAN outperforms the state-of-the-art methods.

\subsubsection{Acknowledgment.}
This research is supported in part by the 2030 National Key AI Program of China 2018AAA0100503 (2018AAA0100500), National Science Foundation of China (No. 61772341, No. 61472254), Shanghai Municipal Science and Technology Commission (No. 18511103002, No. 19510760500, and No. 19511101500), the Program for Changjiang Young Scholars in University of China, the Program for China Top Young Talents, the Program for Shanghai Top Young Talents, Shanghai Engineering Research Center of Digital Education Equipment, and SJTU Global Strategic Partnership Fund (2019 SJTU-HKUST).
\bibliographystyle{splncs04}
\bibliography{my.bib}
%




\end{document}